\documentclass[aps,prd,
superscriptaddress,
showpacs,
preprintnumbers,
nofootinbib,
bibnotes,
amsmath,
amssymb,
twocolumn,
floatfix
]{revtex4-2}
\usepackage[utf8]{inputenc}
\usepackage{array,bbold,booktabs}
\usepackage{todonotes}
\setuptodonotes{inline}
\usepackage[normalem]{ulem}
\newcolumntype{L}[1]{>{\raggedright\let\newline\\\arraybackslash\hspace{0pt}}m{#1}}
\newcolumntype{C}[1]{>{\centering\let\newline\\\arraybackslash\hspace{0pt}}m{#1}}
\newcolumntype{R}[1]{>{\raggedleft\let\newline\\\arraybackslash\hspace{0pt}}m{#1}}

\DeclareMathOperator{\SU}{SU}
\DeclareMathOperator{\End}{End}

\usepackage[caption=false]{subfig}
\usepackage{listings}

\usepackage{xcolor}

\definecolor{codegreen}{rgb}{0,0.6,0}
\definecolor{codegray}{rgb}{0.5,0.5,0.5}
\definecolor{codepurple}{rgb}{0.58,0,0.82}
\definecolor{backcolour}{rgb}{0.95,0.95,0.92}

\lstdefinestyle{mystyle}{
    backgroundcolor=\color{backcolour},   
    commentstyle=\color{codegreen},
    keywordstyle=\color{magenta},
    numberstyle=\tiny\color{codegray},
    stringstyle=\color{codepurple},
    basicstyle=\ttfamily\footnotesize,
    breakatwhitespace=false,         
    breaklines=true,                 
    captionpos=b,                    
    keepspaces=true,                 
    numbers=left,                    
    numbersep=5pt,                  
    showspaces=false,                
    showstringspaces=false,
    showtabs=false,                  
    tabsize=2
}

\usepackage{graphicx}
\usepackage{dcolumn}
\usepackage{bm}
\usepackage{verbatim}
\usepackage{amsfonts}
\usepackage{longtable}
\usepackage{natbib}
\usepackage{hyperref}
\usepackage{dcolumn}
\usepackage[capitalise]{cleveref}
\crefname{section}{Sec.}{Secs.}
\usepackage{xspace}

\newcommand\coarse[1]{\tilde{#1}}
\newcommand\1{\mathbb{1}}

\newcommand\N{\mathbb{N}}

\renewcommand\phi\varphi

\newcommand\regensburg{Department of Physics, University of Regensburg, 93040 Regensburg, Germany}

\begin{document}
\title{Gauge-equivariant pooling layers for preconditioners in lattice QCD}

\author{C.~Lehner}\thanks{Corresponding author}\email{christoph.lehner@ur.de}\affiliation{\regensburg}
\author{T.~Wettig}\affiliation{\regensburg}

\date{\today}

\begin{abstract} 
  We demonstrate that gauge-equivariant pooling and unpooling layers can perform as well as traditional restriction and prolongation layers in multigrid preconditioner models for lattice QCD.  These layers introduce a gauge degree of freedom on the coarse grid, allowing for the use of explicitly gauge-equivariant layers on the coarse grid.  We investigate the construction of coarse-grid gauge fields and study their efficiency in the preconditioner model.  We show that a combined multigrid neural network using a Galerkin construction for the coarse-grid gauge field eliminates critical slowing down.
\end{abstract}

\preprint{}

\keywords{machine learning, lattice QCD}

\maketitle

\section{Introduction}

Numerical simulations of quantum field theories such as Quantum Chromodynamics (QCD) continue to be our best systematically improvable method to obtain information on the nonperturbative features of the theory. These simulations are done on a finite space-time lattice on large supercomputers. In most cases, the run time of the simulations is dominated by the solution of the Dirac equation for a fixed gauge field. This is usually done by iterative algorithms whose iteration count is determined by the condition number of the matrix representing the linear system, in our case the Dirac operator. As we approach the interesting regions of parameter space, i.e., physical quark mass and continuum limit, the condition number of the Dirac matrix becomes very large, leading to critical slowing down. To deal with this problem, very sophisticated algorithms have been developed over the years. In particular, suitably constructed multigrid preconditioners have been shown to reduce or even eliminate critical slowing down \cite{Luscher:2003qa,Luscher:2007se,Brannick:2007ue,Babich:2010qb,Frommer:2013fsa,Boyle:2014rwa,Brannick:2014vda,Yamaguchi:2016kop,Brower:2018ymy,Brower:2020xmc,Boyle:2021wcf}. Multigrid algorithms use restriction and prolongation operators to transfer fields from a fine to a coarse grid and back. In a recent paper \cite{Lehner:2023bba} we discussed the construction of such multigrid preconditioners in the language of gauge-equivariant neural networks and showed that the multigrid paradigm can be learned efficiently by such networks.  However,  in Ref.~\cite{Lehner:2023bba} the restriction and prolongation layers were not learned but computed by hand. The aim of the present paper is to demonstrate that these two layers can also be learned by gauge-equivariant neural networks and perform as well as the traditional construction.  We show that both the model of Ref.~\cite{Lehner:2023bba} as well as the new models discussed in the current work eliminate critical slowing down.

The construction of multigrid algorithms for lattice field theory has a long history, addressing both the Markov chain Monte Carlo sampling of fields and the computation of propagators. In the late 1980s and early 1990s a number of groups devised several multigrid schemes aimed at eliminating critical slowing down for different lattice field theories, gauge groups, and fermion discretizations \cite{Goodman:1986pv,Hulsebos:1988xs,Hulsebos:1989up,Hulsebos:1990er,Hulsebos:1991bx,Vink:1991fa,Hulsebos:1992rn,Ben-Av:1988jfp,Ben-Av:1990lml,Harmatz:1990md,Lauwers:1991zz,Solomon:1991bw,Ben-Av:1993tlk,Harmatz:1992iv,Lauwers:1992cp,Lauwers:1992rd,Brower:1991en,Brower:1991ni,Brower:1990at,Brower:1991xv,Kalkreuter:1990gf,Kalkreuter:1991ir,Kalkreuter:1991xv,Kalkreuter:1992aw,Kalkreuter:1992ba,Kalkreuter:1992ne,Kalkreuter:1993hd}. There was even an early attempt to use neural networks in this context \cite{Baker:1992yv}. These works used gauge-equivariant constructions of restriction and prolongation operators to address high-frequency noise from the gauge degrees of freedom.  Note that in these papers gauge equivariance is referred to as gauge covariance, as is common in quantum field theory.
Another important guiding principle is the approximate preservation of the space spanned by the low eigenmodes of the Dirac operator on the coarse grid. The observation of ``local coherence'' of the low modes \cite{Luscher:2007se} implies that this space can be approximated locally by a relatively small number of suitable vectors. State-of-the-art multigrid algorithms make use of this observation in the construction of the restriction and prolongation operators. Our explicit construction of these operators in \cite{Lehner:2023bba} was also based on this observation. Here, we replace this construction by gauge-equivariant pooling and unpooling layers but are still guided by the same objectives.  These layers are parametrized by gauge-invariant spin matrices which are learned in the present work. In future work, we will construct models that, for a given gauge configuration, provide these matrices as output features.

There is a growing body of related work. Several authors have constructed multigrid algorithms, or elements thereof, using neural networks \cite{Katrutsa:2017,Greenfeld:2019,Luz:2020,Eliasof:2020,Huang:2021,vanBetteray:2022}, but gauge equivariance did not play a role in these papers. Gauge equivariance of neural networks was first discussed in \cite{pmlr-v48-cohenc16,Cohen:2019}. In short, gauge symmetry can be built into the model by requiring that the map implemented by the neural network commutes with local gauge transformations. As a result, the neural network does not need to learn this symmetry and can achieve the same expressivity with fewer weights. Gauge-equivariant neural networks were constructed to generate gauge-field ensembles in several lattice field theories in \cite{Kanwar:2020xzo,Boyda:2020hsi,Abbott:2022zhs}. Neural networks that do not explicitly preserve gauge equivariance were used as preconditioners in a two-dimensional U(1) lattice gauge theory in \cite{Cali:2022qbd}. Reference~\cite{Favoni:2020reg} demonstrated that gauge-equivariant neural networks can approximate any gauge-equivariant function on the lattice. In Ref.~\cite{Aronsson:2023rli} the equivariance of neural networks was extended to global lattice symmetries, and group-equivariant pooling layers were discussed.

This paper is structured as follows.  In Sec.~\ref{sec:coarsening} we describe our coarsening approach using gauge-equivariant pooling and unpooling layers.  In Sec.~\ref{sec:wilson}
we provide details of the Wilson-clover Dirac spectrum on a gauge configuration with nonzero topological charge.  In Sec.~\ref{sec:training} we discuss the training strategy for the (un)pooling layers, and in Sec.~\ref{sec:results} we show that the models resolve critical slowing down.  In Sec.~\ref{sec:summary} we summarize our results and provide an outlook on our future research program.

\section{Gauge-equivariant coarsening}\label{sec:coarsening}

In the following, we build on notation defined in Ref.~\cite{Lehner:2023bba} but introduce an explicitly gauge-equivariant coarsening procedure using gauge-equivariant pooling and unpooling layers that are combined with subsampling layers.

\subsection{Review of notation and coarse-grid vector space}

We consider a $d$-dimensional space-time lattice, the fine grid, and denote the set of its sites by $S$. We define a field $\phi:S \to V_I$, $x \mapsto \phi(x)$ on the fine grid with internal vector space
\begin{align}
    V_I=V_G\otimes V_{\bar G}\,,
\end{align}
where $V_G$ is a gauge vector space and $V_{\bar G}$ is a non-gauge vector space, respectively. The set of such fields is denoted by $\mathcal{F}_\phi$. Under a gauge transformation $\Omega:S \to \End(V_G)$, $x \mapsto \Omega(x)$, the fields transform as $\phi(x) \to \Omega(x) \phi(x)$. Furthermore, we consider gauge fields $U_\mu : S \to \End(V_G)$, $x \mapsto U_\mu(x)$, where $\mu\in\{1,\ldots,d\}$.  In the case of QCD, $U_\mu(x) \in \SU(3) \subset \End(V_G)$.  We will use $U$ as a short-hand notation for the tuple $(U_1,\ldots,U_d)$.

We also consider a $d$-dimensional coarse grid with set of sites $\coarse S$.  We define fields on the coarse grid $\coarse\phi:\coarse{S}\to \coarse V_I$, $y\mapsto \coarse\phi(y)$ with internal vector space $\coarse V_I$.   The set of such coarse fields is denoted by ${\cal F}_{\coarse{\phi}}$.  In contrast to Ref.~\cite{Lehner:2023bba}
\begin{align}
  \coarse V_I=V_G\otimes\coarse V_{\bar G}
\end{align}
i.e., in the current work the local gauge space on the coarse grid is the same as on the fine grid.

As in Ref.~\cite{Lehner:2023bba}, we define a block map $B:\coarse{S} \to {\cal P}(S)$, where $\mathcal{P}$ denotes the power set.  We also define a map 
\begin{align}\label{eqn:refmap}
    B_r: \coarse S \to S,\, y \mapsto B_r(y)
\end{align} that selects for each site $y$ on the coarse grid a reference site $B_r(y)$ on the fine grid.  In the following, we only consider maps $B_r$ for which $B_r(y) \in B(y)$.
The coarse fields shall transform as 
\begin{align}\label{eqn:coarsegauge}
\coarse\phi(y) \to \coarse{\Omega}(y) \coarse\phi(y)    
\end{align}
with
\begin{align}
\coarse{\Omega}(y)=\Omega(B_r(y))
\end{align}
under gauge transformations $\Omega$. For a related discussion of gauge-equivariant blocking schemes, see, e.g., Ref.~\cite{Ben-Av:1993tlk}.

\subsection{Restriction and prolongation layers}
\label{sec:layers}
\begin{figure}[tb]
    \centering
    \includegraphics[scale=0.4]{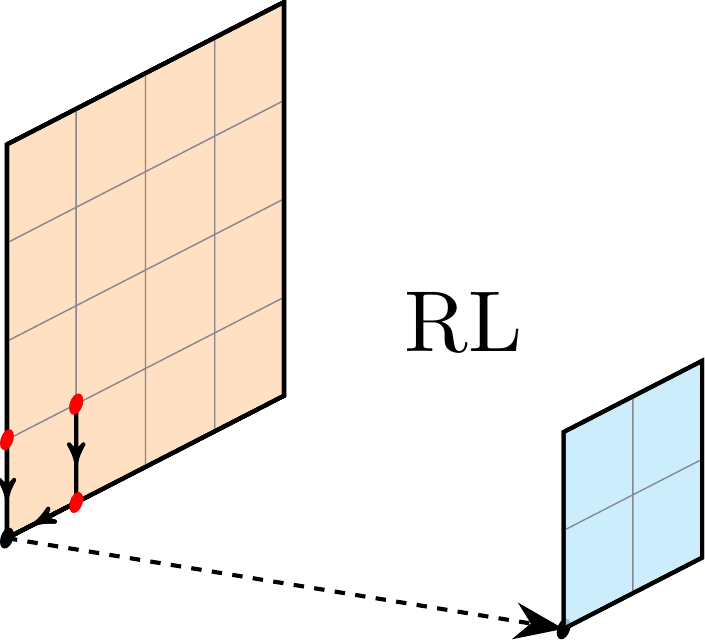}\hspace{15mm}\includegraphics[scale=0.4]{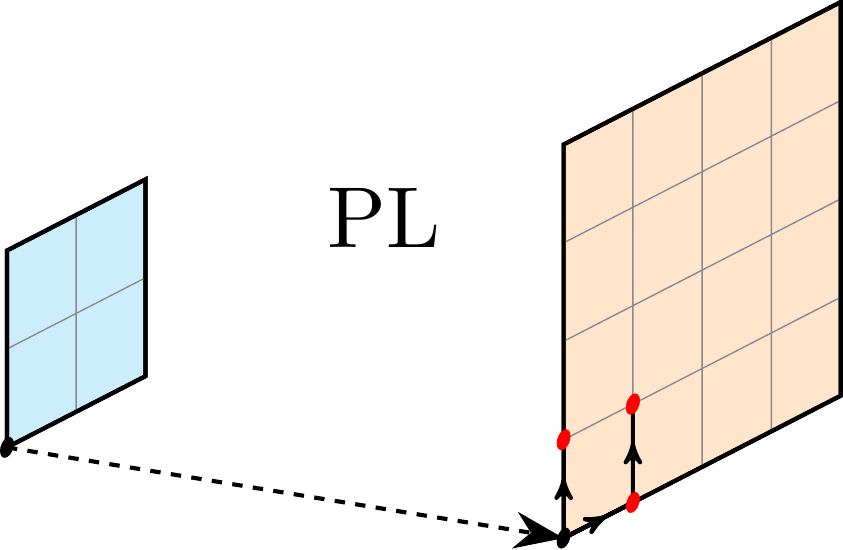}
    \caption{Graphical representation of restriction layer (left) and prolongation layer (right) for a single feature. The input and output features are represented by the planes, and the layers are represented by the paths drawn and the arrow mapping the input to the output feature. The reference site is drawn in black.}
    \label{fig:restrictprolong}
\end{figure}

The restriction layer (RL) can be written as the composition of a pooling layer (Pool) and a subsampling layer (SubSample),
\begin{align}
    \text{RL} = \text{SubSample} \circ \text{Pool} \,.
\end{align}

The pooling layer Pool$:{\cal F}_{\phi} \to {\cal F}_{\phi}$, $\phi \mapsto \text{Pool}\phi$ is given by
\begin{align}\label{eqn:pool}
 \text{Pool}\phi(x) = \sum_{q \in Q} W_q(x) T_q\phi(x) \,.
\end{align}
In the following we describe the elements of this equation in detail. The sum is over couples (i.e., two-tuples) $q=(p,\bar U)$ that consist of a path $p$ and a gauge field $\bar U$.
A path $p$ is defined as a sequence of hops without reference to a starting or end point.  A set of paths $P$ shall be called ``complete'' if it connects every site in $B(y)$ to $B_r(y)$ exactly once.  A complete set of paths therefore always has $\vert B(y) \vert$ elements, where $\vert X\vert$ denotes the cardinality of a set $X$.  In the current work, we only consider couples with $\vert Q \vert = n \vert B(y) \vert$ and $n\in\N^+$ such that $n$ prescriptions to construct the gauge field are combined with $n$ prescriptions to construct a complete set of paths.

The pooling layer is parametrized by weights $W_{q}(x) \in \End(V_{\bar G})$.
In the context of the current paper the $W_q(x)$ are spin matrices.

Finally, the operator $T_q$ for $q=(p,\bar U)$ is the parallel-transport operator $T_p:{\cal F}_\phi\to{\cal F}_\phi$, $\phi\mapsto T_p \phi$ defined in Ref.~\cite{Lehner:2023bba} with gauge fields $U$ replaced by $\bar{U}$.\footnote{In \cref{eqn:pool}, $T_q\phi(x)$ means that the field $T_q\phi$ is evaluated at $x$.}
The gauge fields $\bar{U}$ entering $T_p$ do not have to be the original fine-grid gauge links $U$ as long as they transform in the usual way, i.e., as
\begin{align}
\bar{U}_\mu(x) &\to \Omega(x) \bar{U}_\mu(x) \Omega^\dagger(x+\hat{\mu})
\end{align}
under gauge transformations $\Omega$.  We will make use of this freedom in this work.

\begin{figure*}[t]
    \centering
    \includegraphics[width=\linewidth]{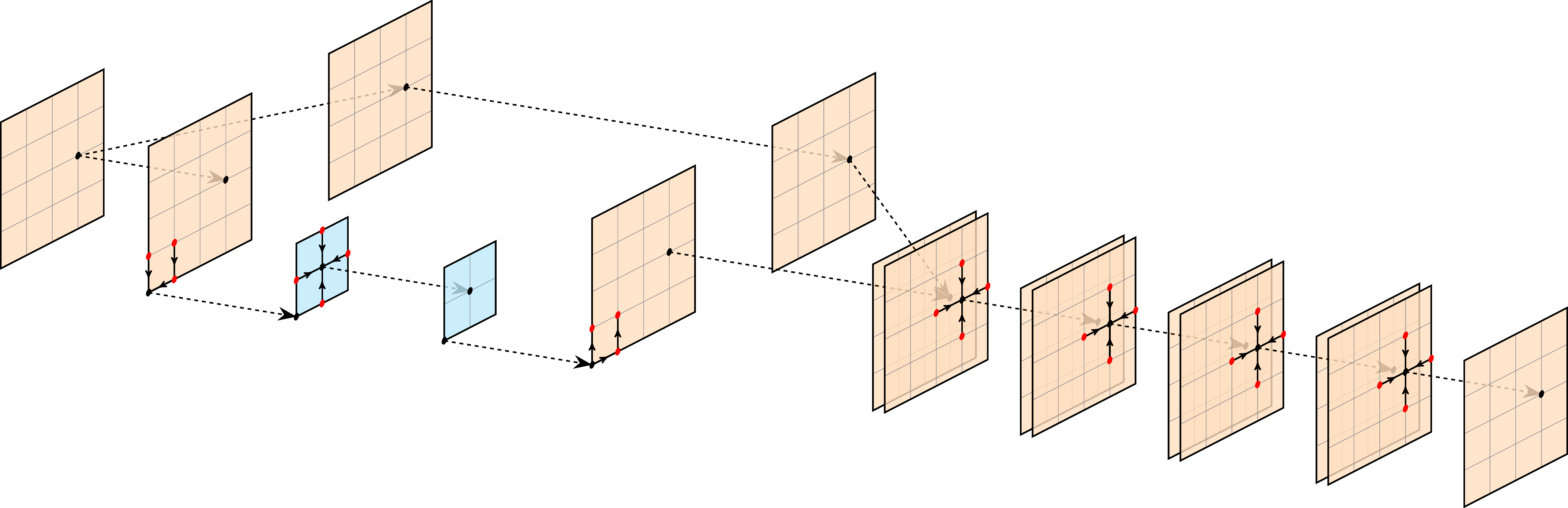}
    \caption{The two-level multigrid model studied in this work.    The model is similar to the one studied in Ref.~\cite{Lehner:2023bba}, but explicitly gauge-equivariant pooling and unpooling layers are used in the current work for the restriction and prolongation layers.  The coarse-grid layer is limited by the blue features.  This layer and the last four layers are LPTC layers introduced in Ref.~\cite{Lehner:2023bba}.}
    \label{fig:multigridmodel}
\end{figure*}

The subsampling layer SubSample$:{\cal F}_{{\phi}}\to{\cal F}_{\coarse{\phi}}$, $\phi \mapsto \text{SubSample}\phi$ is defined by
\begin{align}
  \text{SubSample}\phi(y) = \phi(B_r(y))
\end{align}
for a given choice of reference-point map $B_r$ defined in Eq.~\eqref{eqn:refmap}.  This construction therefore satisfies Eq.~\eqref{eqn:coarsegauge} with $\coarse\phi = \text{RL} \phi$ for a given $\phi \in {\cal F}_\phi$.  For a discussion of a general group-equivariant pooling layer, see Ref.~\cite{Aronsson:2023rli}.

The prolongation layer (PL) is simply defined as
\begin{align}
    \text{PL} = \text{Pool}^\dagger \circ \text{SubSample}^\dagger \,,
\end{align}
where the dagger of an operator $O$ is defined in the usual way by requiring $\phi_1^\dagger O\phi_2=(\phi_2^\dagger O^\dagger\phi_1)^*$ for arbitrary $\phi_1$ and $\phi_2$.
Note that the couples and weights of a restriction and prolongation layer can in principle be chosen independently.  The models studied in this work, however, use the same couples and weights for both RL and PL so that $\text{PL}=\text{RL}^\dagger$.\footnote{In the context of a multigrid solver, Ref.~\cite{Goodman:1989jw} calls this the variational choice because it follows from a variational principle.}

A graphical representation of the restriction and prolongation layers is given in Fig.~\ref{fig:restrictprolong}. The pooling layer is a generalization of the local parallel-transport convolution (LPTC) layer introduced in Ref.~\cite{Lehner:2023bba}. However, one would typically implement the combined RL directly to avoid unnecessary computation of feature elements that will be discarded by the subsequent subsampling layer. 
This can be done efficiently by precomputing, for each complete set of paths, a field $S \to \End(V_G)$ that is used in combination with a reduction operation within each block.
We provide such implementations of both RL and PL in the Grid Python Toolkit (GPT) \cite{GPT}.

We note that the construction of similar restriction and prolongation operations has a long history, see, e.g., \cite{Ben-Av:1988jfp,Ben-Av:1993tlk,Kalkreuter:1991xv}.

\subsection{Coarsening of the gauge fields}

In the current work, we preserve the general model structure introduced in Ref.~\cite{Lehner:2023bba}. However, we replace the restriction and prolongation layers with ones based on gauge-equivariant pooling and unpooling layers, see Fig.~\ref{fig:multigridmodel}. This replacement introduces an explicit gauge degree of freedom on the coarse grid so that the coarse-grid layer can be constructed in an explicitly gauge-equivariant manner.   For this layer we need coarse gauge fields $\coarse{U}$.

The gauge transformation property of coarse fields given in Eq.~\eqref{eqn:coarsegauge} is consistent with gauge fields on the coarse grid that perform a parallel transport between reference sites $B_r(y)$ and $B_r(y^\prime)$ on the fine grid, where $y$ and $y^\prime$ are neighboring sites on the coarse grid.  Such gauge fields must transform as
\begin{align}\label{eqn:coarsegaugetrafo}
    \coarse{U}_\mu(y) &\to \coarse{\Omega}(y) \coarse{U}_\mu(y) \coarse{\Omega}^\dagger(y+\hat{\mu})
\end{align}
under gauge transformations. We investigate two choices for the $\coarse{U}_\mu$ in this work.  

The first choice is to connect $B_r(y)$ and $B_r(y^\prime)$ using the shortest path on the fine grid connecting both points.  In this work, we use a block map $B$ such that $B(y)$ is given by a Cartesian product of neighboring sites in each dimension, and a fixed reference site $B_r$ within each block so that the shortest path is unique and aligns with a coordinate axis.  We then always have
\begin{align}
    B_r(y^\prime) - B_r(y) = b \hat{\mu}
\end{align}
with unit vector $\hat{\mu}$ in direction $\mu$ and $b \in \N^+$.  The coarse-grid gauge field $\coarse{U}_\mu(y)$ corresponding to this pair of reference points is then simply
\begin{align}
 \coarse{U}_\mu(y) = U_\mu(B_r(y)) \cdots  U_\mu(B_r(y) + (b-1)\hat{\mu})
\end{align}
with fine-grid gauge links $U_\mu$.  We will refer to this choice as the ``plain coarse-link model.''

The second choice is based on the Galerkin coarse-grid operator
\begin{align}
    \coarse{D} = \text{RL} \circ D \circ \text{PL}
\end{align}
with gauge-equivariant fine-grid operator $D$.  For the purpose of the current paper, $D$ is the Wilson-clover Dirac operator (for the precise definition see Ref.~\cite{Lehner:2023bba}).  We then simply define
\begin{align}
\label{eq:coarseU}
    \coarse{U}_\mu(y) = \coarse{D}(y, y + \hat{\mu}) \,,
\end{align}
which transforms as in \cref{eqn:coarsegaugetrafo} since
$\coarse{D}(y,y^\prime)$ transforms to $\coarse{\Omega}(y) \coarse{D}(y, y^\prime)\coarse{\Omega}^\dagger(y^\prime)$ under gauge transformations $\Omega$. 
We refer to this choice as the ``Galerkin model.''  Note that in the Galerkin model the coarse gauge links will depend on the weights in the RL and PL.  In the Galerkin model $\tilde{U}_\mu(y) \in \End(\coarse{V}_I)$, while $\tilde{U}_\mu(y) \in \End(V_G)$ in the plain coarse-link model.  Both are acceptable in the context of the gauge-equivariant coarse-grid LPTC layer in Fig.~\ref{fig:multigridmodel} as long as Eq.~\eqref{eqn:coarsegaugetrafo} is satisfied.

We again note that there is a rich history of related work, see, e.g., Refs.~\cite{Nielsen:1981fi,Balaban:1983bj,Mack:1983yi,Ben-Av:1990lml,Kalkreuter:1992ne,Sokal:1993kb}. As in these works, our coarse gauge fields defined by \cref{eq:coarseU} are, in general, no longer elements of the original gauge group. While this is not a problem of principle, Refs.~\cite{Ben-Av:1990lml,Lauwers:1992cp} found better performance of the multigrid algorithm if the coarse gauge fields are projected back to the original gauge group. We plan to investigate this possibility in future work. We also note that there is an alternative way to define the coarse gauge fields using the pooling and subsampling layers introduced in \cref{sec:layers} and applying them to the gauge links between the blocks, see, e.g., Ref.~\cite{Kalkreuter:1992ne}. We did not implement this alternative because it does not increase the expressivity of the model compared to \cref{eq:coarseU}.

\section{Dirac spectrum and topology}\label{sec:wilson}
\begin{figure}[tb]
    \centering
    \includegraphics{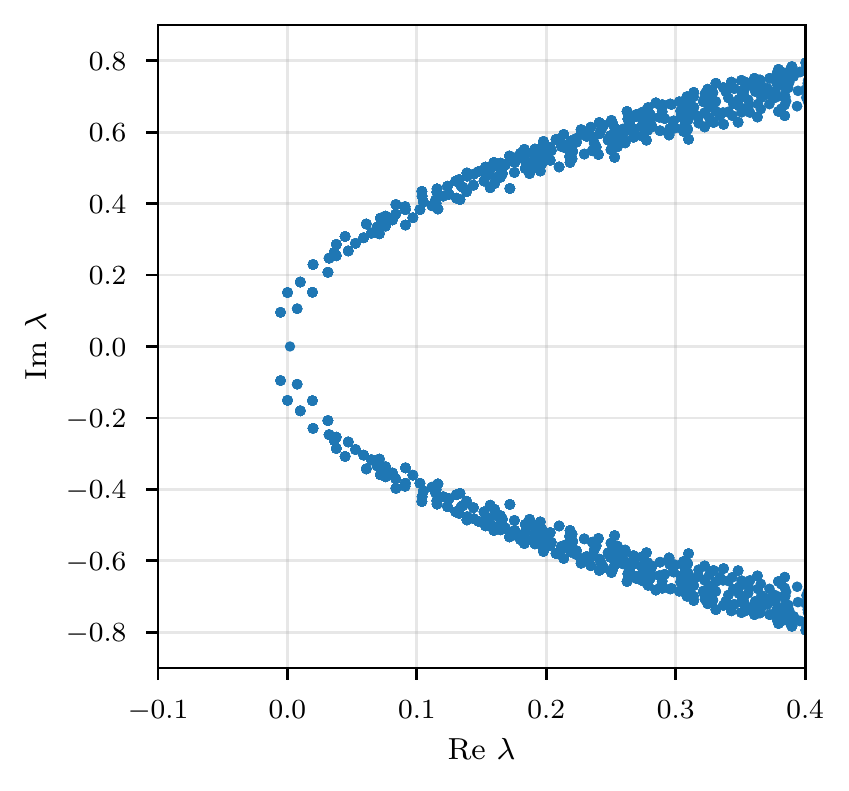}
    \caption{Smallest eigenvalues $\lambda$ of the Wilson-clover Dirac operator with mass $m=-0.5645$ and $c_{\rm sw}=1$ on a pure-Wilson-gauge configuration with topological charge $Q=1$, $\beta=6$, and $8^3 \times 16$ lattice sites.  The mass $m$ is tuned to near criticality for the experiments in this work.}
    \label{fig:spectrum}
\end{figure}

As in Ref.~\cite{Lehner:2023bba}, we have generated quenched Wilson gauge configurations with $8^3 \times 16$ lattice sites for $\beta=6$ and attempt to precondition the Dirac equation for the Wilson-clover Dirac operator with $c_{\rm sw}=1$.  In order to provide an even more challenging setup for the preconditioner models, we select gauge configurations with topological charge $Q=1$  defined via the five-loop enhanced definition of Ref.~\cite{deForcrand:1997esx} after cooling the gauge fields by applying the Wilson flow \cite{Luscher:2010iy} with flow time $t=10$.\footnote{The measured value for the configuration used in this work is $Q=0.998$.}
The Dirac operator has an eigenvalue with vanishing imaginary part and real part very close to the lower edge of the spectrum, see Fig.~\ref{fig:spectrum}.
In this case, we expect critical slowing down to be clearly visible as the quark mass is tuned to criticality.

\section{Training strategy}\label{sec:training}
In the following we describe our training strategy for the preconditioner model shown in Fig.~\ref{fig:multigridmodel}.  We perform the training in two steps.

In the first step, we train only the restriction and prolongation layers.  One may naturally consider to train $\text{PL} \circ \text{RL}$ as an autoencoder with training vectors sampled from the low-mode space of $D$.  We find that this strategy by itself is not sufficient to obtain an efficient model.  Instead, we also train $\text{PL} \circ \text{RL}$ to act as a projector onto the low-mode space, i.e., it should project high modes to zero.  Furthermore, we found that it is beneficial to approximately preserve the property $\text{RL} \circ \text{PL} = \1$.  We also found that restricting $\text{PL}=\text{RL}^\dagger$ by using the same couples $q=(p,\bar U)$ and the same weights $W_q(x)$ for the restriction and prolongation layers did not reduce the performance of the model, and therefore we adopt this choice for simplicity.

We implement this strategy by using the cost function
\begin{align}
    C &= \vert \text{PL} \circ \text{RL} v_\ell - v_\ell \vert^2 + \vert \text{PL} \circ \text{RL} v_h - P_\ell v_h\vert^2 \notag\\
    &\quad + \vert \text{RL} \circ \text{PL} v_c - v_c \vert^2
\end{align}
with two fine-grid vectors $v_\ell$ and $v_h$ and one coarse-grid vector $v_c$.  For each training step new random vectors $v_\ell,v_h,v_c$ are chosen according to the following procedure.  For $v_\ell$ we select a random element of $\{u_1,\ldots,u_s\}$ of the near-null space vectors $u_i$ defined in Ref.~\cite{Lehner:2023bba} with $s \in \N^+$.  For $v_h$ and $v_c$ we take random vectors with elements normally distributed about zero.  The low-mode projector
\begin{align}
  P_\ell = W^\dagger W
\end{align}
with $W$ defined in Eq.~(31) of \cite{Lehner:2023bba}
is using the same set of near-null vectors $\{u_1,\ldots,u_s\}$.
All vectors $v_\ell$, $v_h$, and $v_c$ are normalized to unit length before being used in the cost function.  Note that $P_\ell v_\ell = v_\ell$ by construction so that we can also write the cost function in the more symmetric way
\begin{align}
    C &= \vert \text{PL} \circ \text{RL} v_\ell - P_\ell v_\ell \vert^2 + \vert \text{PL} \circ \text{RL} v_h - P_\ell v_h\vert^2 \notag\\
    &\quad + \vert \text{RL} \circ \text{PL} v_c - v_c \vert^2 \,.
\end{align}

This training procedure provides the gauge-invariant spin matrices $W_q(x)$ for a given gauge configuration.  While the current training strategy does not reduce the overall cost compared to the multigrid model studied in Ref.~\cite{Lehner:2023bba}, we will study constructing the gauge-invariant $W_q(x)$ directly from a given gauge field $U$ using gauge-invariant models \cite{Favoni:2020reg} in future work.  The local features of $W_q(x)$ may be related to features of the local energy density, topological charge density, and general Wilson loops so that no retraining may be needed for a different gauge configuration of the same ensemble.

In the second step, we use the trained RL and PL in the model $M$ of Fig.~\ref{fig:multigridmodel} and train the model with frozen pooling-layer weights using the same cost function as in Ref.~\cite{Lehner:2023bba},
\begin{align}\label{eq:costmg}
 C = \vert M b_h - u_h \vert^2 + \vert M b_\ell - u_\ell \vert^2
\end{align}
with $b_h = D v_1$, $u_h=v_1$, $b_\ell = v_2$, and $u_\ell = D^{-1} v_2$.  Here, $v_1$ and $v_2$ are random vectors normalized such that $\vert b_h \vert = \vert b_\ell \vert = 1$.  After this procedure, we can also continue to train the model without freezing the pooling-layer weights. However, no benefit was observed from this refinement.

\section{Model details and results}\label{sec:results}
In this section we demonstrate the performance of the models we studied with focus on removing the critical slowing down in solving the Dirac equation when the mass parameter $m$ is tuned towards criticality.  

For concreteness, we use a coarse grid of size $2^3 \times 4$ such that $\vert B(y) \vert = 4^4$, and $s=4$.  Note that in Ref.~\cite{Lehner:2023bba} we used $s=12$. However, for the case at hand $s=4$ was sufficient to obtain a well-performing model.  

For the pooling layers, we found that using gauge fields which are smeared differently depending on the set of paths works well.  Concretely, we use 9 different gauge fields $\bar{U}^{(i)}$ with $i=1,\ldots,9$.  We construct the $\bar{U}^{(i)}$ by applying $i (i - 1)/2$ steps of $\rho=0.1$ stout smearing \cite{Morningstar:2003gk} to the unsmeared gauge fields $U$. 
For fixed $i$, we define paths $p^{(ij)}$ that connect all elements of $B(y)$, enumerated by $j=1,\ldots,\vert B(y) \vert$, to the reference site $B_r(y)$.
For different $i$ we use different prescriptions for the paths $p^{(ij)}$, and then use the couples $q_{ij}=(p^{(ij)},\bar{U}^{(i)})$ in \cref{eqn:pool}.
We define four different prescriptions $\hat{p}_{1},\ldots,\hat{p}_{4}$ in the following and set $p^{(ij)}=\hat{p}^{(j)}_{i \mod 4}$.  

For all prescriptions we select the reference site to be the origin of each block.  For the first block the reference site corresponds to  coordinate $(1,1,1,1)$. The starting site for path $p$ is denoted by $(x_1+1,x_2+1,x_3+1,x_4+1)$.  Then
the first prescription to construct the paths is to use $T_p =H_{-4}^{x_4} H_{-3}^{x_3} H_{-2}^{x_2} H_{-1}^{x_1}$
with hopping operator $H_{\pm\mu}$ defined in Ref.~\cite{Lehner:2023bba}.
The second prescription is $T_p = H_{-1}^{x_1}H_{-2}^{x_2} H_{-3}^{x_3} H_{-4}^{x_4}$.  The third and fourth prescriptions modify the first and second prescription, respectively, by permuting the hops in a way that, to the degree possible, at most one hop in one direction is performed at a time.  A concrete example for $x_1=2$, $x_2=2$, $x_3=1$, $x_4=1$ is
$T_p=H_{-2} H_{-1} H_{-4} H_{-3} H_{-2} H_{-1}$ for the third prescription and $T_p=H_{-1} H_{-2} H_{-1} H_{-2} H_{-3} H_{-4} $ for the fourth prescription.  

We investigated many additional choices for reference sites, paths, and gauge fields to be used in the pooling-layer construction. However, the setup just described proved to perform well while still being relatively simple.

\begin{figure}[tb]
    \centering
    \includegraphics{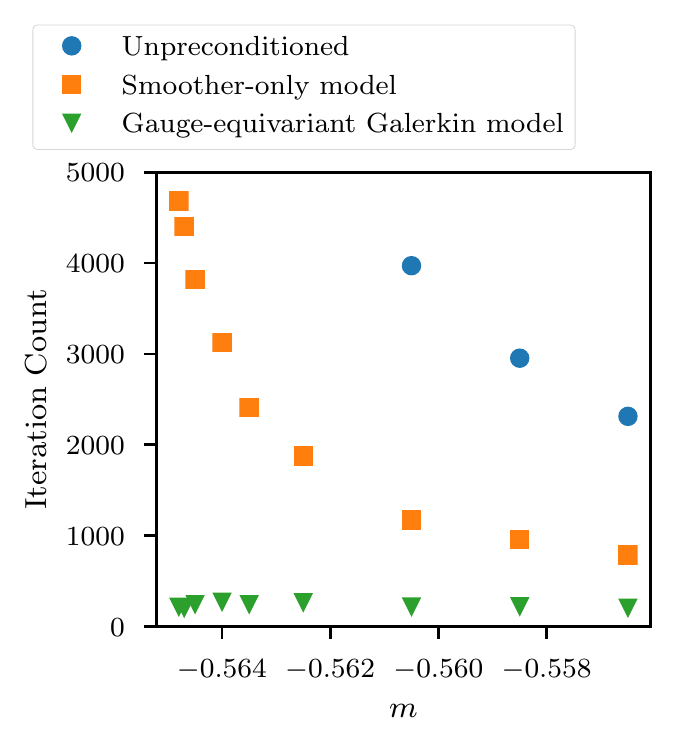}
    \caption{Outer iteration count of unpreconditioned and preconditioned solvers as a function of the quark mass.  The gauge-equivariant Galerkin model completely removes the critical slowing down as the mass is tuned to criticality.}
    \label{fig:criticalslowing1}
\end{figure}

\begin{figure}[tb]
    \centering
    \includegraphics{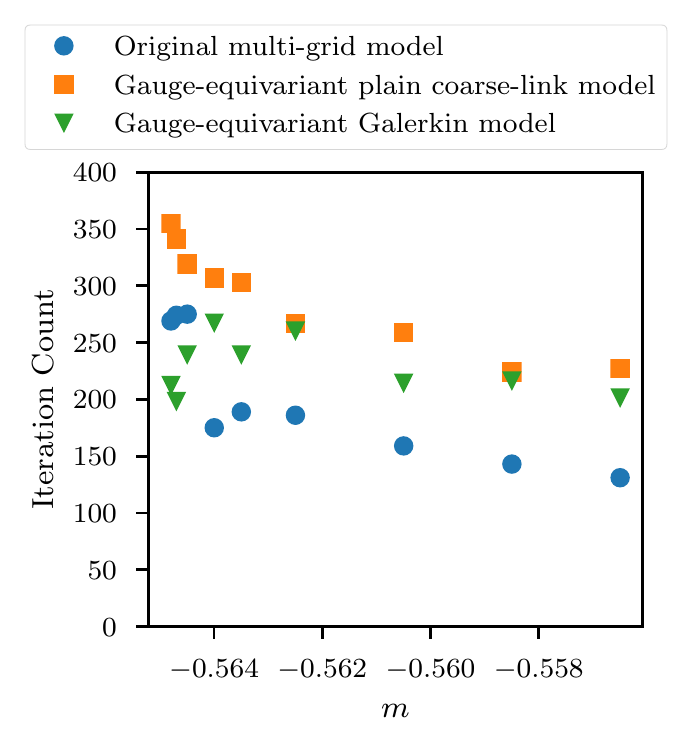}
    \caption{Comparison of multigrid models studied in this work and the original multigrid model of Ref.~\cite{Lehner:2023bba}.  The gauge-equivariant Galerkin model performs very well even for masses near criticality.  The plain coarse-link model shows a mild increase in outer iteration count near criticality.}
    \label{fig:criticalslowing2}
\end{figure}

We then solve the Dirac equation with and without preconditioning and study the iteration count of the outer FGMRES \cite{FGMRES} solver to $10^{-8}$ precision as a function of the quark mass $m$.  In Fig.~\ref{fig:criticalslowing1} we compare the outer iteration count of the unpreconditioned solver with the smoother-only model of Ref.~\cite{Lehner:2023bba} and the new gauge-equivariant Galerkin model.  We find that in the smoother-only model critical slowing down is still visible, while it is completely absent in the gauge-equivariant Galerkin model.  In Fig.~\ref{fig:criticalslowing2} we compare the original multigrid model of Ref.~\cite{Lehner:2023bba} with the gauge-equivariant Galerkin model and with the gauge-equivariant plain coarse-link model.  We find that the original model and the gauge-equivariant Galerkin model perform best, while the plain coarse-link model indicates a small remaining signature of critical slowing down.  Note that there is some randomness in the training procedure that explains the performance fluctuations between neighboring mass points for a given model.

\section{Summary and Outlook}\label{sec:summary}
The current work is part of a larger research program based on gauge-equivariant multigrid neural networks.  In our first paper \cite{Lehner:2023bba} we demonstrated that a state-of-the-art multigrid preconditioner can be learned efficiently by gauge-equivariant neural networks.  The restriction and prolongation layers of Ref.~\cite{Lehner:2023bba} were, however, manually constructed by traditional methods to find near-null-space vectors.  

In the current work, we replaced this construction by gauge-equivariant pooling and unpooling layers that are learned for a given gauge configuration.  We demonstrated that such models can eliminate critical slowing down and perform as well as traditional multigrid models.  The pooling and unpooling layers are parametrized by gauge-invariant spin matrices, which in turn can be learned by models such as discussed in Ref.~\cite{Favoni:2020reg}.  
The construction of such models, including a detailed study of transfer learning, is left for future work.  If successful, such models promise to drastically reduce the setup cost in multigrid preconditioners and may therefore play an important role in improving the performance of gauge-generation algorithms such as HMC \cite{Duane:1987de} or flow-based models \cite{Kanwar:2020xzo,Boyda:2020hsi,Abbott:2022zhs,Abbott:2022zsh,Abbott:2022hkm}.

Another important topic left for future studies is the construction of multigrid models for operators with a more challenging spectrum, such as domain-wall fermions.  

Finally, gauge-equivariant multigrid models should also be able to learn to directly approximate complex hadronic correlation functions without constructing them from intermediate approximations of propagators.  Such direct approximations can then be used to reduce statistical noise without introducing bias \cite{Shintani:2014vja}.  We will explore this application of gauge-equivariant multigrid models in future work as well.

\bigskip

{\bf Acknowledgments.}
This work was funded in part by the Deutsche Forschungsgemeinschaft (DFG, German Research Foundation), project number 460248186.

\bibliography{references}

\end{document}